\documentclass[aps,prb,twocolumn,groupedaddress]{revtex4}
\usepackage{graphicx}
\usepackage{amsmath}
\usepackage{natbib}
\begin{document}

\title{Bulk measurement of copper and sodium content in CuIn$_{0.7}$Ga$_{0.3}$Se$_2$ (CIGS) solar cells with nanosecond pulse length laser induced breakdown spectroscopy (LIBS)}
\author{Jeremy M. D. Kowalczyk}
\affiliation{Department of Physics and Astronomy, University of Hawai`i M$\bar{a}$noa, Honolulu, HI 96822}
\author{Jeffrey J. Perkins}
\affiliation{Department of Physics and Astronomy, University of Hawai`i M$\bar{a}$noa, Honolulu, HI 96822}
\author{Alexander DeAngelis}
\affiliation{Hawai`i Natural Energy Institute, University of Hawai`i M$\bar{a}$noa, Honolulu, HI 96822}
\author{Jess Kaneshiro}
\affiliation{Hawai`i Natural Energy Institute, University of Hawai`i M$\bar{a}$noa, Honolulu, HI 96822}
\author{Stewart A. Mallory}
\affiliation{Hawai`i Natural Energy Institute, University of Hawai`i M$\bar{a}$noa, Honolulu, HI 96822}
\author{Yuancheng Chang}
\affiliation{Hawai`i Natural Energy Institute, University of Hawai`i M$\bar{a}$noa, Honolulu, HI 96822}
\author{Nicolas Gaillard}
\affiliation{Hawai`i Natural Energy Institute, University of Hawai`i M$\bar{a}$noa, Honolulu, HI 96822}

\date{\today}
\begin{abstract}
In this work, we show that laser induced breakdown spectroscopy (LIBS) with a nanosecond pulse laser can be used to measure the copper and sodium content of CuIn$_{0.7}$Ga$_{0.3}$Se$_2$ thin film solar cells on molybdenum.  This method has four significant advantages over methods currently being employed: the method is inexpensive, measurements can be taken in times on the order of one second, without high vacuum, and at distances up to 5 meters or more.  The final two points allow for in-line monitoring of device fabrication in laboratory or industrial environments.  Specifically, we report a linear relationship between the copper and sodium spectral lines from LIBS and the atomic fraction of copper and sodium measured via secondary ion mass spectroscopy (SIMS), discuss the ablation process of this material with a nanosecond pulse laser compared to shorter pulse duration lasers, and examine the depth resolution of nanosecond pulse LIBS.
\end{abstract}
\maketitle

\section{Introduction}
Laser induced breakdown spectroscopy (LIBS) has long been used as a method for detection of trace elements \cite{Miziolek2006} and has more recently been shown to be an effective method for measuring atomic fraction of constituents with depth \cite{vadillo1997,Anderson1995,Kim2012}.  We have shown previously that a linear relationship between the sodium peak from LIBS and the mass of sodium deposited during deposition exists \cite{Kowalczyk2010}; in this study, we build on that work and use LIBS to measure the atomic fraction of copper and sodium in CIGS solar cells via calibration with secondary ion mass spectroscopy (SIMS).  We specifically focused on sodium, the control of which in CIGS solar cells is critical to achieve high efficiency devices \cite{Braunger2000}, but to date composition measurement techniques either cannot accurately quantify the sub-one atomic percent levels (for example, x-ray fluorescence (XRF) and energy dispersive X-ray spectroscopy (EDX)) or are expensive, time consuming, require high vacuum, and cannot be done at large distances (examples include EDX, SIMS, and x-ray photo-electron spectroscopy (XPS)).  LIBS is an inexpensive and fast method that can be done at a large range of sample to instrument distances (from centimeters to meters) without the need for vacuum and is able to detect and quantify copper and sodium atomic percentages.  We know of three other studies have been done on CIGS with LIBS. Pilkington showed that LIBS could produce meaningful depth profiles of constituents of CIGS \cite{Pilkington2006}. Lee showed a linear relationship between Ga concentration and the Ga/In and Ga/Cu ratios \cite{Lee2012}. Kim showed that sodium produces a strong LIBS signal \cite{Kim2012}. Here we calibrate LIBS to quantify copper and sodium content over a range of different copper and sodium concentrations at a sample to detector distance of 5.2 meters.  Due to the limited bandwidth of the spectrometer available to us (383 nm to 900 nm), and lower efficiency of our LIBS setup in the blue/UV end of the spectrum (and subsequently lower signal to noise ratio (SNR)), we were unable to measure atomic lines for gallium, indium, and selenium (though measurement of selenium is difficult for others reasons discussed in the depth profile discussion below).

\section{Theory}
\subsection{LIBS spectra theory}
LIBS employs a high power laser to create a plasma from the material to be investigated \cite{Winefordner2004}.  As the plasma cools and ions and electrons recombine, photons characteristic of the transition energies of the elements present are emitted, and these spectral lines are then matched to those of known lines of specific elements.  In the most simple cases, the number of measured counts of an atom's spectral line is proportional to the atomic percent as has been shown, for example, for chromium in steels \cite{Stipe2010}.  This technique damages an area of the target on the order of the spot size of the laser.

In general, the nature of the matrix of other materials in which a particular element resides will influence the ratio of the atomic percent of that element to the intensity of its corresponding lines (the so called 'matrix effect') \cite{Clegg2009}.  It has been shown that in some cases this ratio can be constant even over broad compositional ranges \cite{Perkins2009}. In this study, the concentration range of interest of all constituents is narrow, so we assumed and confirmed in our results a linear relationship between the relative intensity of the atomic lines of copper and sodium and their atomic concentrations.

\subsection{LIBS plasma formation and ablation}
\begin{figure}[htp]
\centering
\includegraphics[scale=1.0,angle=0]{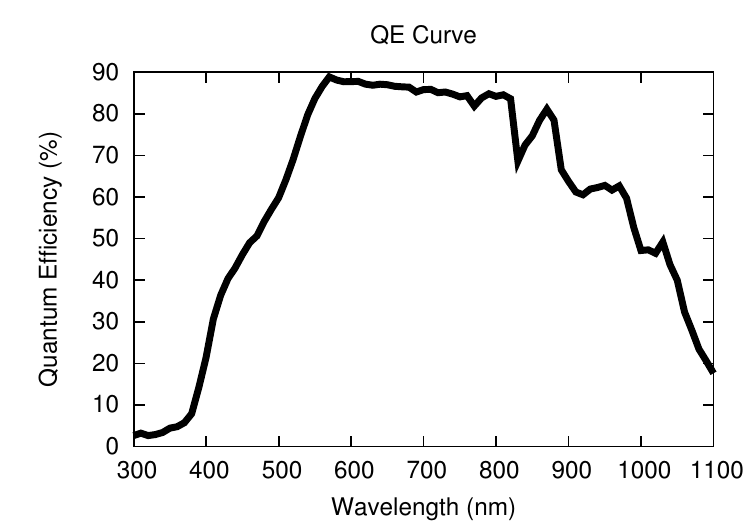}
\caption[Typical quantum efficiency curve for the devices used in this study]{Typical quantum efficiency curve for the devices used in this study.  This curves show the conversion efficiencies of photons at a particular wavelength to photocurrent production.  Note the significant quantum efficiency for photon wavelengths less than 1100 nm (energy greater than 1.13 eV) which confirms that the band gap is below the YAG laser photon energy of 1.17 eV.}
\label{fig:QE}
\end{figure}

In semiconductors like CIGS there are few conduction band electrons and the initial laser radiation for LIBS leads to either single or multi-photon band to band transitions instead of direct heating of the material \cite{Winefordner2004}.  After generation of conduction band electrons, the lattice heats due to coupling of the generated electron-hole pairs to the lattice. The photon absorption cross section is roughly 17 orders of magnitude greater for photons with energy above the band gap energy compared to below \cite{Miller1994}, leading to a higher density of deposited energy with photon energy above the band gap. The band gap of the CIGS materials created at the Hawai`i Natural Energy Institute and used for this study has been measured to be $\sim1.2$ eV, corroborated by the significant photon to electron-hole pair conversion evident from the quantum efficiency (QE) curves in figure \ref{fig:QE} at wavelengths shorter than 1100 nm (energy greater than 1.13 eV).  The laser wavelength used for LIBS in this study was 1064 nm (photon energy of 1.17 eV), enabling laser energy to be densely coupled into the CIGS material.

If the pulse duration is short compared to the lattice ion relaxation time (on the order of 100 fs to 1 ps), the initially excited electron-hole pairs do not have time to couple to the lattice before all of the pulse energy is deposited \cite{Chichkov1996}.  Since the interaction cross section of photons with the lattice is small compared to that of photons to the electron-hole excitation, this leads to a dense deposition of energy in the material that is typically greater than its heat vaporization, resulting in sublimation of the material, no mixing of the material at different depths, and an abrupt edge to the ablation crater.

Conversely, if the laser pulse duration is longer than the picosecond time scale, significant energy is coupled into the lattice phonon modes and diffuses into the material before the threshold energy density for vaporization can be reached.  When this occurs, the material is first melted, then the melted material is vaporized.  This sequence results in mixing of material at different depths and a rough edge to the ablation crater \cite{Chichkov1996}.  The nanosecond pulses used in this work are much longer than the lattice ion relaxation time and result in a rough ablation edge as will be discussed in the results section.

\section{Experimental}
\subsection{CIGS PV cell fabrication}
\begin{figure}[htp]
\centering
\includegraphics[scale=1.5,angle=0]{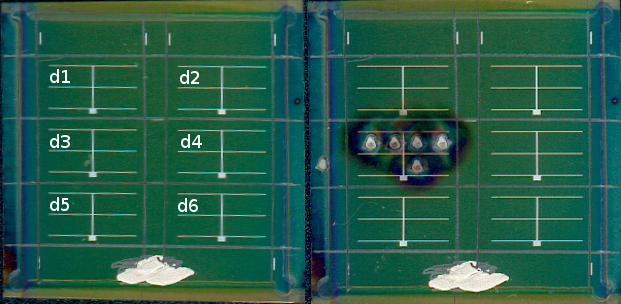}
\caption[Image of one substrate with 6 devices]{Image of one substrate with 6 devices (labeled d1 through d6) before LIBS (left) and after LIBS (right).  Note the five craters formed from the five measurements taken on device 3 (d3).  Five such devices spanning 3 substrates were used for this study.}
\label{fig:PV_cell}
\end{figure}

For this work a total of 18 PV cells were fabricated on 3 substrates in the following stack: a 1"x1" thin titanium substrate, 1 $\mu m$ of molybdenum deposited via magnetron sputtering, 10 mg of NaF deposited via evaporation, 1.8 $\mu m$ of CIGS deposited via a 3-stage evaporation process \cite{Contreras1994}, 100 nm of CdS n-type buffer layer deposited via chemical bath deposition, 80 nm of ZnO, 100 nm of indium tin oxide (ITO), and Ni/Ag grids for charge collection.  The complete cells are pictured on the left in figure \ref{fig:PV_cell}.  During the CIGS deposition, the evaporation sources for copper and sodium were intentionally positioned to provide a significant gradient of evaporant across the substrates to be analyzed.  This non-optimal fabrication process resulted in functional solar cells with modest performance between 7 and 9 percent efficiency but allowed for a range of copper and sodium contents with which our LIBS signals could be calibrated.

\subsection{LIBS setup}
\begin{figure}[htp]
\centering
\includegraphics[scale=0.45,angle=0]{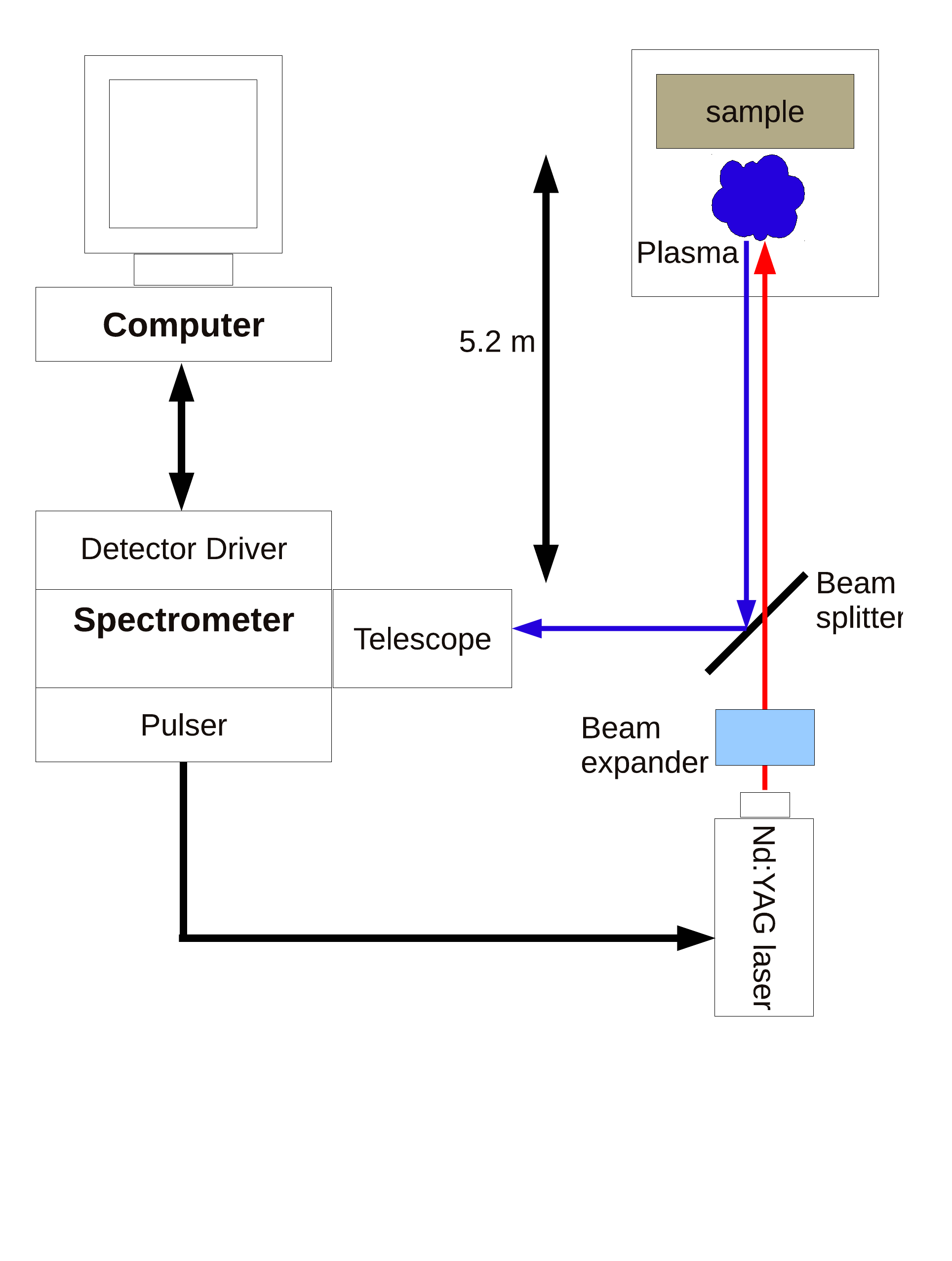}
\caption[Schematic of the LIBS measurement setup]{Schematic of the LIBS measurement setup.  Laser travels through the beamsplitter, is focused on the sample by a lens, and causes plasma formation of the sample.  The sample emits photons which travel down to the beam splitter and are reflected into the telescope for collection by the spectrometer.  A computer controls the spectrometer integration window and laser pulse timing.}
\label{fig:LIBS_cartoon}
\end{figure}

Figure \ref{fig:LIBS_cartoon} shows a schematic of the LIBS measurement setup.  The samples were interrogated in air with a 1064 nm Nd:YAG Continuum Surelite II laser at 100 mJ per pulse with a 600 $\mu m$ spot (fluence of 35 J/cm$^2$) with an integration time of 1 second (note we did not have access to an ICCD camera for this study, so the integration time was long leading to higher noise levels and the need for higher power pulses to increase the SNR).  This relatively high laser pulse energy was chosen to yield an SNR ratio of 10:1 for all measurements.  A near-infrared (NIR) dichroic beam-splitter was used to assure collinearity between the laser and telescope optical axes. The resultant radiation was fiber coupled to an Ocean Optics LIBS 2500+ Spectrometer with a usable bandwidth of 500 nm to 900 nm.  The results were digitally recorded.  The current experiment is designed for detection at a distance of $5.2$ m from the samples to emulate an in-line device fabrication setup.

An optically transparent fused silica window with a 1064 nm transmission of greater than 90 percent was used to allow transmission of the laser and resulting breakdown radiation.  For purposes of system detection calibration, a calibrated tungsten filament thermal source was used to determine the system efficiency as a function of wavelength.

Five spots on each device were analyzed using LIBS (see figure \ref{fig:PV_cell}) and the remaining area of the device was analyzed with SIMS.  Twenty five 100 mJ pulses were used to collect spectra down to $\sim 3.4 \mu m$ (measured with a Tencor Alpha-Step profilometer) from the top of each device .  The background in the LIBS spectra (due to Bremsstrahlung radiation, stray light, and detector dark counts) was removed to improve the SNR.

\section{Results and Discussion}
\subsection{Analysis of LIBS craters}
\begin{figure}[htp]
\centering
\includegraphics[scale=0.70,angle=0]{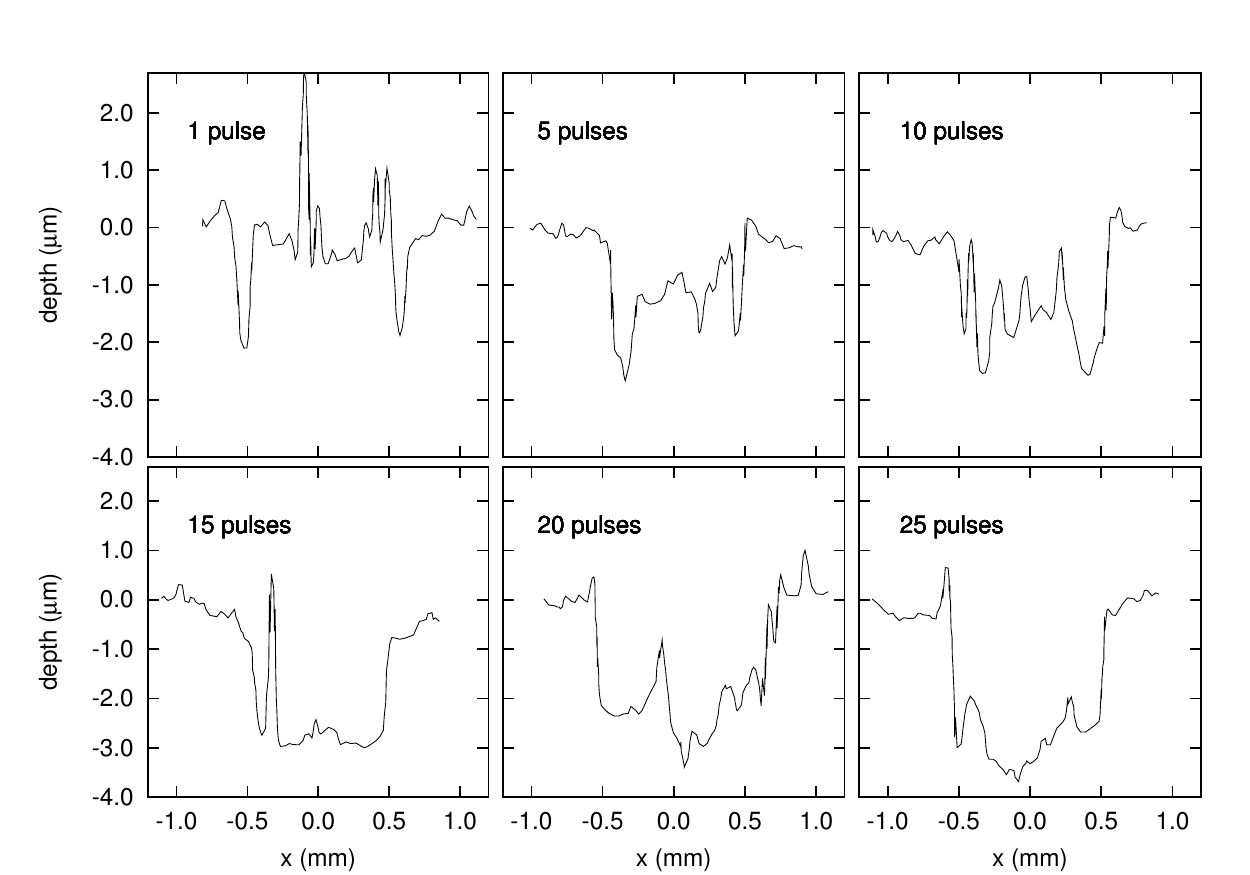}
\caption[Profiles of the craters from LIBS as a function of pulse number]{Profiles of the craters from LIBS as a function of pulse number. Here x is the distance transverse to the incoming laser beam. Note the roughness induced by the long pulse length of the laser.}
\label{fig:profilometer}
\end{figure}

Profiles of the LIBS craters are presented in figure \ref{fig:profilometer}.  These show a relatively flat floor to the craters in some cases, but a rough floor in others.  We assume the roughness is due to the long (nanosecond) pulses leading to melting, evaporation, and ultimate resolidification of the melted material under the influence of chaotic forces from the plasma into irregular shapes.  The uneven ablation for each pulse leaves the utility of nanosecond LIBS in questions for fine depth analysis, but has little effect on the bulk analysis done here since the variations in the crater floors are small compared to the overall depth analyzed.  

\subsection{LIBS peak identification}
In order to correlate the spectra obtained from LIBS to the presence of particular atoms, strong lines without overlap to other lines in the material that were within the bandwidth of the spectrometer were chosen from the NIST Atomic Spectra Database \cite{nistasdmisc} as shown in table \ref{lines_table}.  Note that with a spectrometer sensitive to the appropriate spectral range, strong lines for indium, gallium (at 410 and 417 nm respectively\cite{Lee2012}), and tin could be recorded allowing LIBS analysis of all constituents of the device except selenium. 

\begin{table}[htp]
\caption{Strong spectral lines and their corresponding elements used for LIBS analysis.}
\centering
\begin{tabular}{| c | c | }
 \hline
  wavelength (nm) & element \\
 \hline
  510.55 & Cu \\ 
  550.65 & Mo \\ 
  589.00 & Na \\ 
  625.81 & Ti \\ 
  636.23 & Zn \\ 
  643.85 & Cd \\ 
 \hline
\end{tabular}
\label{lines_table}
\end{table}

\subsection{Copper and sodium calibration curves}
\begin{figure}[htp]
\centering
\includegraphics[scale=1.0,angle=0]{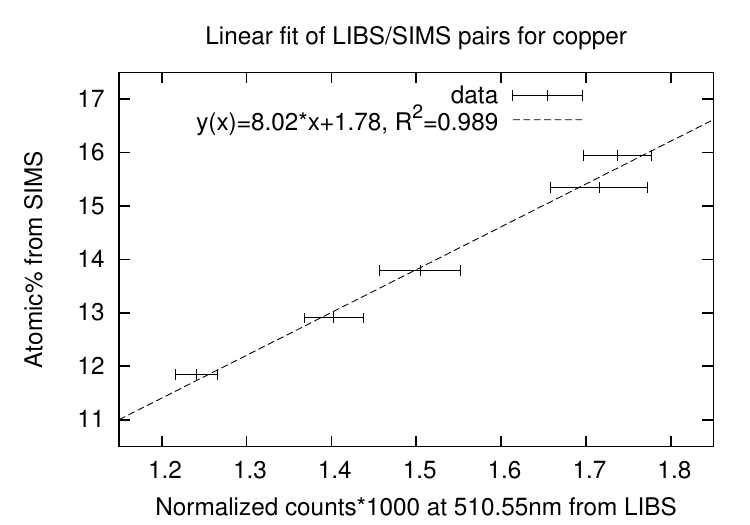}
\caption[Linear fit of atomic fraction of copper]{Linear fit of atomic fraction of copper from SIMS analysis versus normalized photon counts from LIBS at 510.55 nm.  Five LIBS measurements were taken and averaged for each data point.  The error bars represent plus and minus one standard deviation.  Note that the atomic percentage of copper in the CIGS layer is typically 25\%, but here we have taken an average over the entire device through the molybdenum layer making the values on the y-axis lower.} 
\label{fig:cu_lin}
\end{figure}

\begin{figure}[htp]
\centering
\includegraphics[scale=1.0,angle=0]{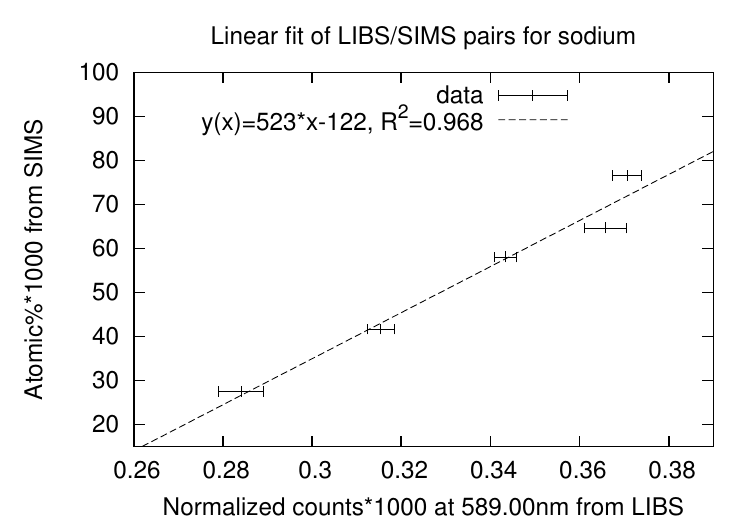}
\caption[Linear fit of atomic fraction of sodium]{Linear fit of atomic fraction of sodium from SIMS analysis versus normalized spectrometer counts from LIBS at 589.00 nm.  Five LIBS measurements were taken and averaged for each data point.  The error bars represent plus and minus one standard deviation. Note this is a measurement of the total amount of sodium in the stack including the molybdenum layer which contains significant amounts of sodium.}
\label{fig:na_lin}
\end{figure}

As can be seen from figure \ref{fig:profilometer} multiple pulses were required to obtain spectra for the entire device and therefore to use LIBS to measure its contents.  However, due to saturation of the spectrometer, multiple spectra needed to be taken at each spot.  In order to obtain an accurate calibration of the LIBS spectra to the atomic percent, all of the spectra originating from the ablated material (3.4 $\mu m$ of device material from the 25 pulses) in a particular spot were added together and the sum was normalized to the total number of spectrometer counts.  This procedure yields a normalized spectrum representative of the bulk of the device for a particular spot.  LIBS signals taken from the 5 spots on each of 5 devices (see figure \ref{fig:PV_cell}) were averaged and the standard deviation was calculated.   SIMS analysis was done on each of the five devices and the atomic percent was averaged over the same 3.4 $\mu m$ of ablated material as the LIBS analysis.  The SIMS bulk average atomic percent (y coordinate) and LIBS bulk average peak height (x coordinate) data pairs for the copper and sodium are plotted along with a linear least squares fit in figures \ref{fig:cu_lin} and \ref{fig:na_lin}.  The fits for both copper and sodium are good with values for $R^2=0.989$ and $R^2=0.968$, respectively, showing that indeed that LIBS produces a reliable signal proportional to the atomic fraction of copper and sodium in this concentration regime.  Note that the matrix of other materials that the copper and sodium reside in changes with each pulse (as shown in figure \ref{fig:depth_LIBS}) so some pulses may have a different matrix effect induced change in ratio of LIBS counts to atomic percent as a function of depth that we did not explore.  This possible variation is washed out by averaging of all of the 25 pulses and is reduced by the melting and mixing that homogenizes the sample.  Despite this variation, a linear calibration of copper and sodium content in the entire stack was achieved with nanosecond LIBS.

\subsection{Depth profile discussion}
\begin{figure}[htp]
\centering
\includegraphics[scale=1.0,angle=0]{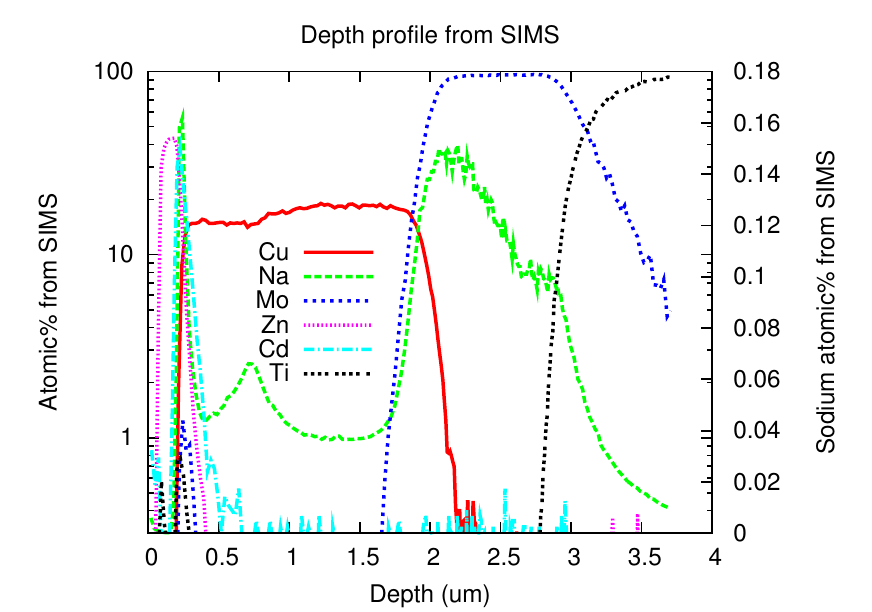}
\caption[SIMS analysis of the median sodium and copper content device.]{SIMS analysis of the median sodium and copper content device.  One can clearly see the different device layers from the signal of their constituent elements.  The ZnO is evident from the zinc signal near the surface.  Below, the cadmium from the CdS layer is visible.  Next the CIGS layer is indicated by the presence of copper.  Finally the molybdenum and titanium substrates are indicated by increased signal for those elements.}
\label{fig:depth_SIMS}
\end{figure}

\begin{figure}[htp]
\centering
\includegraphics[scale=1.0,angle=0]{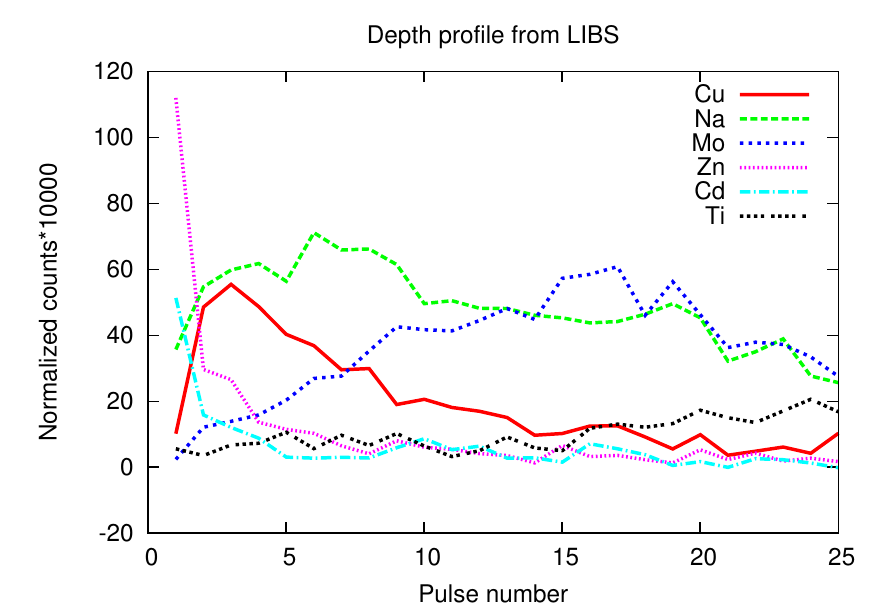}
\caption[Intensity of atomic lines in table \ref{lines_table} as a function of pulse number]{Intensity (normalized to the total spectrometer counts) of atomic lines in table \ref{lines_table} as a function of pulse number.}
\label{fig:depth_LIBS}
\end{figure}

Recently there have been a number of studies \cite{vadillo1997,Anderson1995,Pilkington2006,Papazoglou2004} of LIBS as a depth profiling method.  They have found that generally picosecond or femtosecond pulses are required to achieve sublimation and uniform ablation instead of the melting/evaporation and non-uniform ablation seen with the nanosecond pulses used in this study.  Due to the mixing of layers with depth during melting and the subsequently non-uniform surface created for the next ablating pulse, the abrupt interfaces between layers seen in SIMS analysis in figure \ref{fig:depth_SIMS} are not observed in the corresponding LIBS analysis in figure \ref{fig:depth_LIBS}.

However, the LIBS depth analysis does have the expected qualitative features.  Spectra from only the first three pulses contain zinc and cadmium peaks, as expected from the ZnO and CdS layers in the top 300 nm of the device.  The copper peak slowly decreases until the entire CIGS layer has been ablated by pulse 25.  Finally, the molybdenum and sodium peaks both decrease as the titanium signal is increasing at pulse 25 where SIMS analysis indicated the molybdenum/titanium interface to be.

There are some interesting features of the LIBS depth analysis in figure \ref{fig:depth_LIBS}.  The copper peak is greatest in the first couple of pulses and decreases linearly.  This is most likely due to the ratio of CIGS to molybdenum in the melted material decreasing with depth (corroborated by the increasing molybdenum signal in this range).  Additionally, the molybdenum peak appears surprisingly early in the laser pulses (around pulse 3).  This can be explained by the deep probing of the first few laser pulses around the edge of the crater evident in figure \ref{fig:profilometer}a.  From this phenomenon, we theorize that the entire CIGS layer is melted during the laser pulse and resolidifies between pulses.  This makes LIBS with nanosecond pulses impractical for analysis of selenium in CIGS materials as there is significant evaporation of selenium in the melted material. At the deposition temperature for the CIGS devices of 600 $^o$C, selenium evaporation is significant enough to require a constant vapor pressure of selenium in the deposition chamber in order to maintain stoichiometry.  Assuming the CIGS layer totally melts and its melting point is near that of copper (1085 $^o$C), a liquefied pool of CIGS will evaporate significant selenium with each pulse (and subsequent melting), making accurate measurement of selenium concentration impossible with the technique used here. 

The SIMS analysis in figure \ref{fig:depth_SIMS} shows that there is significant copper and sodium in the molybdenum layer and figure \ref{fig:depth_LIBS} shows that the molybdenum layer is probed long before the entire CIGS layer has been ablated in our LIBS analysis.  This means that our measurement of copper and sodium is the total amount in the CIGS plus molybdenum back contact stack, not just in the CIGS absorber layer.
 
\section{Conclusion}
We have shown that a bulk compositional analysis of the copper and sodium content in a complete CIGS solar cell is possible with LIBS; calibration curves were obtained showing that the atomic percentage is linearly proportional to the peak heights of the corresponding element in the compositional range for functional CIGS devices.  Additionally, while we have shown nanosecond LIBS is not a viable method for precise depth profiling of this material at the fluence used in this study.  The viability of LIBS for bulk analysis of CIGS solar cells and other thin film technology demonstrated here can lead to significant cost and time savings when compared to methods currently in use.  

\section{Acknowledgments}
Fabrication of the solar cells used in this study was funded by the United States Air Force Research Laboratory Space Vehicles Directorate as part of the Rapidly Deployable Solar Electricity and Fuel Sources program.  Special thanks goes to Shiv Sharma and David Bates at the Hawai`i Institute for Geophysics for generously providing time on their LIBS setup, background removal of the LIBS data, and their input and feedback.  Finally, we would like to thank the Honolulu chapter of the Achievement Rewards for College Scientists (ARCS) Foundation for granting one of the authors (JMDK) the Doris and Robert Pulley Award which was used to pay for the SIMS analysis.

\bibliography{cigs}

\end{document}